\documentstyle[prd,aps,floats,epsfig]{revtex}
\tighten
\begin{document} 
\draft
\preprint{
\begin{tabular}{rr}
CfPA-97-TH-31\\
IC-96/97-39\\
\end{tabular}
}

\title{The closet non-Gaussianity of anisotropic Gaussian fluctuations}
\author{Pedro G. Ferreira$^{1}$ and Jo\~ao Magueijo$^2$}
\address{$^{(1)}$Center for Particle Astrophysics,
University of California, Berkeley  CA 94720-7304,USA\\
$^{(2)}$The Blackett Laboratory, Imperial College,
Prince Consort Road, London SW7 2BZ, UK}
\maketitle
\begin{abstract}
In this paper we explore the connection between anisotropic Gaussian 
fluctuations and isotropic non-Gaussian fluctuations.
We first set up a large angle framework for characterizing
non-Gaussian fluctuations: large angle non-Gaussian spectra. 
We then consider anisotropic Gaussian fluctuations in two different
situations. Firstly we look at anisotropic space-times and propose
a prescription for superimposed Gaussian fluctuations; we
argue against accidental symmetry in the fluctuations and that
therefore the fluctuations should be anisotropic. We 
show how these fluctuations display previously known non-Gaussian effects 
both in the angular power spectrum and in non-Gaussian spectra. 
Secondly we consider the anisotropic Grischuk-Zel'dovich effect. 
We construct a flat space time with anisotropic, non-trivial topology and
show how Gaussian fluctuations in such a space-time look non-Gaussian.
In particular we show how non-Gaussian spectra may probe superhorizon
anisotropy. 
\end{abstract}

\date{\today}

\begin{picture}(0,0)
\put(410,240){
\begin{tabular}{rr}
CfPA-97-TH-\\
IC-96/97-39\\
\end{tabular}
}
\end{picture} \vspace*{-0.4 in}

\renewcommand{\thefootnote}{\arabic{footnote}}
\setcounter{footnote}{0}
\section{Introduction}

Anisotropic models of the Universe have been often considered in the past
(eg. \cite{bianchi,krasinski}). In recent times globally anisotropic spacetimes
have attracted attention
for their thought provoking value, as primordial anisotropy would
appear to contradict inflation \cite{baPrinc}. 
It is therefore important to find
experimental evidence for, or constraints on, primordial anisotropy.
The cosmic microwave background  (CMB) is the cleanest
and most accurate experimental probe in current cosmology. Thus
it makes sense to explore the impact of anisotropic expansion on the CMB.
For homogeneous space times this
 was largely done in \cite{colhawk,BJS,bfs}. In the more 
sophisticated analysis in \cite{bfs} the effects
of the unperturbed anisotropic expansion were combined with a
spectrum of superposed Gaussian fluctuations. An admitted shortcoming 
of this analysis is the assumption that while the unperturbed model leaves an
anisotropic pattern in the sky, the Gaussian fluctuations around it are
isotropic. Should the Gaussian fluctuations in such models be anisotropic  
one may expect a more stringent statistical bound on anisotropy, 
if the Universe is indeed isotropic.
One can consider another class of models where the background
space-time is homogeneous and isotropic but anisotropic topological
identifications lead to anisotropic Gaussian fluctuations. Some of
these universes have been considered before \cite{CSS} and an example of 
the patterns in an open universe has been presented in
\cite{LBBS}.

The apparently unrelated issue of large-angle CMB non-Gaussianity 
has also been considered recently, both as an experimental matter 
\cite{kogut}, 
and as a possible prediction in topological defect theories 
\cite{conf,ltx,ltx1,nt,nt1,nt2,sh}. 
In \cite{conf}, in particular, an outline 
is given of a comprehensive formalism for encoding large angle non-Gaussianity
based on the spherical harmonic coefficients $a^\ell_m$ in the expansion
\begin{equation}\label{alm}
\frac{\Delta T(\bf{n})}{T}   
= \sum_{\ell=0}^{\infty} \sum_{m=-\ell}^{\ell}
a^{\ell}_ m Y^{\ell}_m (\bf{n})
\end{equation}
In \cite{conf} it is also stated that ``any non-Gaussian theory
is to some extent anisotropic, favouring particular directions
in the sky and some $m$'s over others''.
The converse statement follows: that Gaussian anisotropic fluctuations
will appear as non-Gaussian fluctuations from the standpoint of
an isotropic theory. This establishes an interesting link between 
the search for cosmological anisotropy and the search for non-Gaussian
signatures. 

Let us consider
Gaussian theories which favour an axis, where $\Omega$ are angles defining
this axis. Then the probability distribution conditional to this axis 
$P(a^\ell_m|\Omega)$ is Gaussian. Isotropy is violated, but the resulting
theory is Gaussian within the reduced set of symmetries the theory now must
satisfy. However from an isotropic point of
view the full ensemble is made up of all the ensembles which favour
an axis, but allowing the axis to be uniformly distributed. 
Such a super-ensemble would undoubtedly be isotropic, but 
it would also be non-Gaussian. Marginalizing with respect to the axis
reveals a non-Gaussian theory, that is
\begin{eqnarray}
P(a^\ell_m)&=&\int d\Omega P(\Omega)P(a^\ell_m|z_1)\nonumber\\
P(\Omega)&=&{\sin\theta\over 4\pi}
\end{eqnarray}
is non-Gaussian. This identifies the origin of the Gaussian/non-Gaussian
switch. Conditionalizing to an axis
renders the theory Gaussian (and anisotropic). Marginalizing with respect to the axis
reveals a non-Gaussian theory (but an isotropic ensemble).

This phenomenon turns out to be a particular case of the general
phenomenon discussed in connection with the 
texture analytical model in \cite{ltx,ltx1}. In that model it is
found that the temperature anisotropies
are very non-Gaussian. The theory has $C_\ell$ cosmic variance error
bars above their Gaussian value, and there are strong correlations among $C_\ell$. 
It turns out, however, that these large-angle 
non-Gaussian effects are largely due to the last texture (as in the texture
closest to us, or the texture at lower redshift). The culprit
identified, one then notices that conditionalizing the theory
to the last texture redshift $z_1$ reveals a Gaussian ensemble, that is,
the probability distribution $P(a^\ell_m|z_1)$ is Gaussian.
Marginalizing with respect to $z_1$, however, produces a non-Gaussian ensemble,
that is the probability
\begin{equation}
P(a^\ell_m)=\int dz_1 P(z_1)P(a^\ell_m|z_1)
\end{equation}
is non-Gaussian.

The picture is then clear\cite{note}. We come up with a construction where the
full ensemble is made up of sub-ensembles which are Gaussian. Each
sub-ensemble is however labelled by an index which from the
the point of view of the full ensemble is a random variable.  
Marginalizing with respect to this variable reveals a non-Gaussian
ensemble. Conditionalizing with respect to this index renders
the theory Gaussian. Such an index was called in \cite{ltx1}
the random index, and it was conjectured 
\footnote{This conjecture can in fact be promoted to a mathematical 
theorem; see~\cite{charlo}.} in that paper that 
non-Gaussianity could often be characterized by a set of such
indices labelling Gaussian ensembles. Within such a construction the 
strategy for predicting experiment must be modified. 
One should now not provide a direct statistical description of the full
ensemble (that is, marginal distributions),
which would be plagued by all sorts of non-Gaussian effects.
Rather it makes more sense to supply information on all the Gaussian sub-ensembles,
plus the distribution function of their random indices.

Hence we may use a sub-class of the comprehensive formalism
for encoding large-angle non-Gaussianity outlined in \cite{conf}
to describe anisotropic Gaussian fluctuations. This is essentially
a large-angle generalization of \cite{fermag} and is described 
in Section~\ref{lang}. The idea is to
complement the angular power spectrum $C_\ell$ with a set of
multipole shape spectra $B_m$ describing how the power is distributed
among the $m$'s for a given scale $l$.  The $B_m$ encode information
on the shape of large angle structures. They are uniformly
distributed in a Gaussian isotropic theory, meaning its
fluctuations are shapeless. However,  as we shall see in Section~\ref{nganisogauss}, 
preferred shapes emerge in non-Gaussian isotropic theories, 
as well as in Gaussian anisotropic
theories, where the $B_m$ are not uniformly distributed. Non-Gaussian spectra
then appear as a natural predictive tool for these theories.

In this paper we study the disguised non-Gaussianity of anisotropic
Gaussian fluctuations along two lines. Firstly, in Section~\ref{glob1},  
we propose a simple method for defining anisotropic Gaussian 
fluctuations. Breaking isotropy essentially amounts to choosing an alternative 
symmetry group under which the covariance matrix should be invariant,
and which picks a favoured direction in the sky. We can then write down
the most general form for the covariance matrix of the theory 
simply by studying the representation theory of the symmetry group.
We argue that the accidental symmetry allowing anisotropic fluctuations 
to be isotropic is a model dependent and unnatural assumption. Hence
Gaussian fluctuations in anisotropic Universes should be anisotropic too.
Although we concentrate on anisotropic fluctuations
with an $SO(2)$ symmetry, the definition and considerations given
in Section II are quite general, as explained in more detail in
Appendix~I.

We then show how anisotropic Gaussian
theories induce well known non-Gaussian effects in the relation between
the observed and the predicted angular power spectrum $C_\ell$.
These effects include larger cosmic variance error bars, and
also the phenomenon of cosmic covariance, that is correlations between 
the observed $C_\ell$. Cosmic covariance allows for more structure 
to exist in each realization than in the predicted average power spectrum
and complicates comparison between theory and experiment. These
effects are shown to be present for anisotropic Gaussian theories in
Section~\ref{power}.

Then, in Section~\ref{nganisogauss}, we show how anisotropic Gaussian
fluctuations render non-Gaussian spectra non-uniformly distributed,
as announced above. We also find the most general class of isotropic 
non-Gaussian theories into which anisotropic Gaussian fluctuations may be mapped.
As a concrete example in Section~\ref{glob2} we proceed
to characterize the non-Gaussian spectra for the relevant, globally
anisotropic
space times.

Along a totally different line in Section~\ref{gz} we construct a simple 
example of a topologically
non-trivial space time and show how the non-Gaussian spectra will
indicate anisotropic topological identifications. We propose this
as an {\it anisotropic} Grischuk-Zel'Dovich effect: from subhorizon,
large angle observables we can characterize super-horizon anisotropies.

In Section~\ref{conc} we discuss the implications of our results and
their practical implementation.
\section{Large-angle non-Gaussianity}\label{lang}


We now set up a formalism for describing large-angle non-Gaussianity
which is based on
\cite{fermag}, but makes use of $a^\ell_m$ coefficients rather than Fourier
components, and so is suitable for mapping large-angle non-Gaussianity.
Again the idea is to map the $\{a^\ell_m\}$ into a set of spectra which for
a Gaussian isotropic theory are independent random variables. 
One of these spectra 
is the angular power spectrum $C_\ell$, and should be a $\chi^2_{2\ell+1}$
for a Gaussian isotropic theory. The other variables make up non-Gaussian
spectra which should be uniformly distributed for a Gaussian isotropic
theory.

The transformation proposed is defined as follows. Firstly
we split the complex modes into moduli and phases
\begin{eqnarray}
a^\ell_0&=&s^l_0\rho^\ell_0\nonumber \\
a^\ell_m&=&{\rho^\ell_m\over {\sqrt 2}} e^{i\phi^\ell_m}
\end{eqnarray}
where $s^l_0=\pm 1$ is simply the sign of $a^\ell_0$.
The fact that the $m=0$ mode is real introduces
a slight modification to the construction in \cite{fermag}.
There are now $\ell+1$ moduli, but there are only $\ell$ phases
(the index $m$ starts at 1 for the phases). Working out the Jacobian
of the transformation shows that for a Gaussian theory
the distribution of the $\{\rho^\ell_m,\phi^\ell_m, s^\ell_0 \}$ is
\begin{equation}
F(\rho^\ell_m,\phi^\ell_m,s^\ell_0)={2{\exp{\left(-{{\sum_0^\ell}\rho_m^2
\over 2C_\ell}\right)}}\over (2\pi)^{1/2}C_\ell^{\ell+1/2}}
\times{\left({\prod_1^\ell
\rho_m} \right)}\times{1\over (2\pi)^\ell}\times {1\over 2}
\end{equation}
The phases $\phi^\ell_m$ are uniformly distributed in $[0,2\pi]$. 
The sign $s^\ell_0$ has a uniform discrete distribution.
The moduli
$\rho^\ell_m$ are $\chi^2_2$ distributed except for $\rho_0^\ell$ 
which is $\chi^2_1$
distributed. Since $\rho_0$ now does not appear in the Jacobian
of the transformation, the only way one can proceed with the construction
in \cite{fermag} is by ordering the $\rho$'s by decreasing order
of $m$, and then introduce polars:
\begin{eqnarray}
\rho_\ell^\ell&=&r\cos\theta_1\nonumber\\
\rho_{\ell-1}^\ell&=&r\sin\theta_1\cos\theta_2\nonumber\\
&...&\nonumber\\
\rho^\ell_1&=&r\sin\theta_1...\cos\theta_l\nonumber\\
\rho^\ell_0&=&r\sin\theta_1...\sin\theta_l
\end{eqnarray}
Again, working out the jacobian of the transformation implies that
for a Gaussian isotropic theory the distribution of these variables
is
\begin{equation}
F(r,\theta_m,s^\ell_0,\phi_m)=
{{\exp {\left(-r^2/ 2C_\ell\right)}}r^{2\ell}\over 
(\pi/2)^{1/2}C_\ell^{\ell+1/2}}
{\prod_1^\ell}\cos\theta_m (\sin\theta_m)^{2(\ell-i)}
\times {1\over 2}\times {1\over (2\pi)^\ell}
\end{equation}
One can then define shape spectra $B^\ell_m$ as
\begin{equation}
B^\ell_m=(\sin\theta_m)^{2(\ell-m)+1}
\end{equation}
so that for a Gaussian isotropic theory one has:
\begin{equation}
F(r,B^\ell_m,s^\ell_0,\phi^\ell_m)=
{{\exp{\left( -r^2 /2 C_\ell\right)}}r^{2\ell}\over 
(\pi/2)^{1/2}C_\ell^{\ell+1/2}(2\ell-1)!!}\times {1\over 2}
\times {1\over (2\pi)^\ell}
\end{equation}
The angular power spectrum $C^\ell$ seen as a random variable is then
related to $r$ by
\begin{equation}
C^\ell={r^2\over 2\ell+1}
\end{equation}
and is a $\chi^2_{2\ell+1}$. The multipole shape spectra $B^\ell_m$
may be obtained from the moduli $\rho^\ell_m$ according to
\begin{equation}
B^\ell_m={\left(\rho^{\ell 2}_{m-1}+\cdots+\rho^{\ell 2}_0
\over \rho^{\ell 2}_{m}+\cdots+\rho^{\ell 2}_0\right)}^{\ell -m +1/2}
\end{equation}
and are uniformly distributed in $[0,1]$. Finally the phases $\phi^\ell_m$
are uniformly distributed in $[0,2\pi]$, and the sign $s^\ell_0$ is
a discrete uniform distribution over $\{-1,+1\}$. 

As in \cite{fermag} we define non-Gaussian structure in terms of
departures from uniformity and independence in the 
$\{B^\ell_m,\phi^\ell_m\}$. Gaussian theories 
can only allow for modulation, that is, a non-constant power spectrum.
The most general power spectrum has as much information as Gaussian
theories can carry. White-noise is the only type of fluctuations
which is more limited in terms of structure than Gaussian 
fluctuations
\footnote{It is curious to note that white noise has less structure
than generic Gaussian fluctuations, but it also has more symmetry.
It is tempting to associate reduction of symmetry and addition
of structure. Anisotropic fluctuations have less symmetry than isotropic
fluctuations, but they also have more structure, reflected in their
non-Gaussian structure.}.
In isotropic Gaussian theories there is no structure in the $\{B^\ell_m,\phi^\ell_m\}$
since these are independent and uniformly distributed.
By allowing the $B^\ell_m$ to be not uniformly distributed, or to 
be constrained by correlations amongst themselves and with the
power spectrum, one adds shape to the multipoles. This is because the
$B^\ell_m$ tell us how the power in multipole $\ell$ given by
$C^\ell$ (or $r$) is distributed among the various $|m|$ modes,
which reflect the shape of the fluctuations.
Indeed the $m=0$ mode (zonal mode) has
no azimuthal dependence. It corresponds to fluctuations with 
strict cylindrical symmetry (rather than statistical symmetry).
The $|m|>0$ modes correspond to the various azimuthal frequencies
allowed for the scale $\ell$. Each of these modes represent a way
in which strict cylindric symmetry may be broken.  The relative
intensities of all the $m$ modes carry information on the shape of
the random structures at least as seen by the scale $\ell$. 
In a Gaussian theory all the $m$ modes must have the same intensity,
something which can be rephrased by the statement that the $B^\ell_m$
are independent and uniformly distributed. Hence Gaussian fluctuation
display shapeless multipoles. Any departure from this distribution
in the $B^\ell_m$ may then be regarded as a evidence for more or less random
shape in the fluctuations.

On the other hand the phases $\phi^\ell_m$ transform under azimuthal
rotations. Therefore they carry information  on the localization of the 
fluctuations. If the phases
are independent and uniformly distributed then the perturbations
are delocalized.

Finally there may be correlations between the various scales defined
by $\ell$. In the language of \cite{fermag} this is what is called
connectivity of the fluctuations. These correlations measure how much
coherent interference is allowed between different scales, a phenomenon
required for the rather abstract shapes and localization on each scale
to become something visually recognizable as shapeful or localized.
As in \cite{fermag} this may be cast into inter-$\ell$ correlators.
As we shall see these are in fact quite complicated for general anisotropic
Gaussian theories. Therefore we have chosen not to dwell on this aspect
of large-scale non-Gaussianity in this paper.

\section{A possible method for introducing Gaussian Fluctuations 
in Anisotropic Universes}
\label{glob1}

We now present a possible way of introducing Gaussian
fluctuations in anisotropic Universes such as the Bianchi models. 
In Section~\ref{gz} we will present another context in which
anisotropy appears: periodic Universes. There we shall present more specific
calculations of anisotropic Gaussian perturbations. Here we shall however
use a method which relies simply on inspecting the reduced symmetry 
group anisotropic Gaussian perturbations must satisfy.
This is a simple, if somewhat phenomenological, way of introducing
the most general Gaussian perturbation which can live in an 
anisotropic background. Without
actually performing a detailed perturbation analysis of these
spacetimes, one can refine the analysis of \cite{bfs} by using this
prescription and possibly find more stringent constraints. 

Let an all-sky temperature anisotropy map be decomposed into
spherical harmonics as in Eqn.~(\ref{alm}).
Then, for a general Gaussian theory, the $a^\ell_m$ are Gaussian 
random variables specified by a covariance matrix which must 
satisfy the symmetries of the underlying theory. In Friedman
models the symmetry group is $SO(3)$, but the symmetry group may be
smaller.
Anisotropic Gaussian fluctuations may be defined 
as Gaussian fluctuations with a covariance matrix satisfying a 
symmetry group which picks a favoured direction in the sky.
We concentrate on anisotropic fluctuations with an $SO(2)$
symmetry, that is, with cylindrical symmetry.

The general form of the covariance matrix may be obtained just from
the representation theory of the symmetry group. The symmetry group breaks the 
$\{a^\ell_m\}$ space into irreducible representations (Irreps). 
The $a^\ell_m$ may then be reexpressed in a basis adapted to these Irreps.
Using Schur's Lemmas \cite{cornwell} one knows (see Appendix~I for more
detail) that the covariance 
matrix of the theory must be a multiple of the identity within each Irrep
\footnote{Schurs' Lemma only applies to finite dimensional
representations, such as the ones offered by the $a^\ell_m$.
If one instead looks at the real space maps $\delta T/T$,
then the representation space is $S^2$. This is infinite dimensional,
and indeed the covariance matrix of Gaussian theories is not diagonal,
and is specified by the two-point correlation function $C(\theta)$. 
}. Furthermore correlations between different $a^\ell_m$ can only
occur for elements of different but equivalent Irreps. 
Hence, for any Gaussian theory subject to a symmetry which does not
lead to equivalent Irreps, the spherical harmonic coefficients, expressed
in a basis adapted to the partition into Irreps, must be 
independent random variables, and their variance must be a function
only of the Irrep they belong to. As we shall see it may happen that
the variance is the same for a set of Irreps. This degeneracy then
leads to an accidental enlarged symmetry. If some of the Irreps are
equivalent then in principle one may also have correlations between
coefficients belonging to different but equivalent Irreps.

As an example consider an isotropic theory. Then the $\{a^\ell_m\}$
for each $\ell$ are an Irrep of the symmetry group $SO(3)$ represented 
by the $D$ matrices
\begin{equation}
R(\psi,\theta,\phi)a^\ell_m=D^\ell_{mm'}(\psi,\theta,\phi)a^\ell_{m'}
\end{equation}
where $(\psi,\theta,\phi)$ are Euler angles. None of these Irreps is equivalent,
as indeed none of them have the same dimension.
Hence for a Gaussian isotropic theory the $a^\ell_m$ must
have a covariance matrix of the form
\begin{equation}
{\langle a^\ell_m a^{\ell' \star}_{m'}\rangle}
=\delta_{\ell \ell'}\delta_{m m'}C_\ell
\end{equation}
If the angular power spectrum $C_\ell$ happens to be a constant
(white-noise) over a certain section of the spectrum then this
degeneracy increases the symmetry group of the theory:
rotations among different $\ell$'s are now an extra symmetry.
This is an accidental symmetry resulting from the degeneracy
displayed by the particular model considered (white noise)
and not required by the underlying theory.

Now suppose that the symmetry group is $SO(2)$, that is, the
unperturbed model supporting the fluctuations is cylindrically
symmetric. Then there is a favoured axis in the Universe and
with respect to this axis the symmetry transformations are
\begin{equation}
R(\phi) a^\ell_m=e^{im\phi}a^\ell_m
\end{equation}
The Irreps are now indexed by $\ell,m$ with $m\ge 0$. They are
one  dimensional complex Irreps for $m> 0$, and one dimensional real
(and trivial) Irreps for $m=0$. For the same $m$ Irreps with
different $\ell$ are equivalent Irreps.
For each $\ell$ we have a single Irrep of $SO(3)$ which splits
into $\ell+1$ Irreps of $SO(2)$. The covariance matrix of the
theory now has the general form:
\begin{equation}\label{ani2p}
{\langle a^\ell_m a^{\ell' \star}_{m'}\rangle}
=\delta_{m m'}C^{\ell\ell'}_{|m|}
\end{equation}
and we may call the diagonal terms $C_{\ell m}$ of $C^{\ell\ell'}_{|m|}$
the cylindrical power spectrum. It may now happen that $C^{\ell\ell'}_{|m|}
=\delta^{\ell\ell'}C^{\ell}_{ |m|}$, and furthermore that a given model displays
the degeneracy $C_{\ell |m|}=C_\ell$, that is the cylindrical power
spectrum is white noise in $m$. In this case the $SO(3)$
symmetry is accidentally restored. However this is no different 
from the white-noise model $C_\ell={\rm const}$
referred to above. It is merely
an accidental enlarged symmetry displayed by a concrete model
and not a fundamental symmetry imposed by the underlying model. 

Accidental symmetries (eg. family symmetry in particle
physics)  are always regarded with horror. If they happen to exist, 
sooner or later a fundamental principle is sought which will promote 
them from accidental to fundamental symmetries. If they don't happen
to exist a priori, such as in the case of fluctuations in anisotropic 
models, then better not postulate them in the first place.

\section{Non-Gaussian effects on the angular power spectrum}\label{power}
Gaussian anisotropic theories display many of the novelties
present in non-Gaussian theories, such as the texture models
considered in \cite{ltx,ltx1}. They trade their added predictivity
in terms of non-Gaussian spectra for larger cosmic variance error
bars in the angular power spectrum. Also the observed $C_\ell$
may be correlated, a phenomenon called cosmic covariance and
present in the texture models in \cite{ltx,ltx1}. Cosmic covariance
(or $C^\ell$ aliasing) induces great mess when comparing predicted
and observed power spectra. Correlations allow for each observed
power spectrum to have more structure than the average power spectrum.
This may result in the average power spectrum corresponding to nothing
that any observer ever sees. More subtle methods for predicting
power spectra are then necessary. Two prescriptions are given in \cite{ltx1}.

\subsection{Cosmic variance surplus}
For a Gaussian isotropic theory the angular power spectrum 
\begin{equation}
C^\ell={1\over 2\ell+1}{\sum^\ell_{m=-\ell}}|a^\ell_m|^2
\end{equation}
has the variance
\begin{equation}
\sigma^2(C^\ell)={2C_\ell^2\over 2\ell+1}
\end{equation}
Here we use the notation $C^\ell$ to denote the random variable and $C_\ell$
to denote its ensemble average.
For a Gaussian anisotropic theory this variance is
\begin{equation}
\sigma^2(C^\ell)={2\over (2\ell+1)^2}{\sum^\ell_{m=-\ell}}C_{\ell m}^2
\end{equation}
If we define the average cylindrical power spectrum by
\begin{equation}
C_\ell={1\over 2\ell+1}{\sum^\ell_{m=-\ell}} C_{\ell m}
\end{equation}
then 
\begin{equation}
\sigma^2(C^\ell)\geq {2C_\ell^2\over 2\ell+1}
\end{equation}
It is a simple analysis exercise to prove this inequality and show that 
it is saturated only when $C_{\ell m}=C_\ell$,
that is when the fluctuations are isotropic.

Generally we may interpret this result as a reduction in the number
of degrees of freedom in the $\chi^2$ induced by anisotropy.
Suppose, for instance, that a theory is strongly anisotropic so
that only a few $m$ modes among the available $2\ell+1$ contribute to the
power spectrum $C_\ell$, for a given $\ell$. 
Then, effectively, the observed power
spectrum $C^\ell$ is the result of these few modes. Since these
are still Gaussian variables the observed power spectrum
is a $\chi^2$, but with an effective number of degrees of freedom
equal to the number of predominant modes. If for example all the
power is concentrate on the $m=0$ mode, then the $C^\ell$ is
a $\chi^2_1$. If all the power is in a $m>0$ mode, the $C^\ell$
is a $\chi^2_2$.

We may use the ratio between the actual cosmic variance of the theory
and its Gaussian prediction to quantify how anisotropic
the fluctuations are. Quantitatively let us call anisotropy 
in the multipole $\ell$ to the quantity
\begin{equation}
A_\ell={\sigma^2_{\rm GA}(C^\ell)\over \sigma^2_{\rm GI}(C^\ell)}=
{1\over 2\ell+1}{\sum^\ell_{m=-\ell}}{\left(C_{\ell m}\over C_\ell
\right)}^2
\end{equation}
which varies between $A_\ell=1$ for isotropic theories to 
$A_\ell=2\ell+1$ for cylindrically symmetric  multipoles (for which
all the power is in the $m=0$ mode).

\subsection{Cosmic covariance}
There are also correlations between different $C^\ell$. For $\ell\ne\ell'$
we have that
\begin{equation}
{\rm cov}(C^\ell,C^{\ell'})={1 \over (2\ell+1)(2\ell'+1)}
{\sum_{m,m'}}{\rm cov}(|a^\ell_m|^2,|a^{\ell'}_{m'}|^2)
\end{equation}
For two
(possibly correlated) complex Gaussian random variables $z_1$ and $z_2$ with
uncorrelated real and imaginary parts, it can be shown that ${\rm
cov}[|z_1|^2,|z_2|^2]={\langle z_1z^*_2\rangle}^2+ {\langle z_1z_2\rangle}^2$,
and so
\begin{equation}\label{covmat}
{\rm cov}(C^\ell,C^{\ell'})={1 \over (2\ell+1)(2\ell'+1)}
{\sum_{m}}C^{\ell \ell' 2}_m 
\end{equation}
where $m$ in the summation runs from  $-{\rm min} (\ell,\ell ')$
to ${\rm min} (\ell,\ell ')$. The off-diagonal elements (in $\ell,\ell'$) in
$C^{\ell\ell'}_m$ therefore induce correlations among the various
observed $C^\ell$. A possible, but model dependent, way to do away
with these correlations is to rotate the $C^\ell$ among themselves
so as to diagonalize
the covariance matrix (\ref{covmat}).  These rotated $C^\ell$ will then
be independent, and so their average value is a good prediction for what 
each observer will see. Also, as shown in \cite{ltx1}, in the rotated basis
the cosmic variance error bars tend to be smaller and approach their Gaussian
minimum. Therefore cosmic covariance, and larger cosmic variance error
bars can be dealt with by means of this trick. However this trick does depend
on each particular model, and is not a Universal prescription applicable
to every model.

\section{The non-Gaussian structures exhibited by anisotropic Gaussian theories}
\label{nganisogauss}
Anisotropic Gaussian theories also display non-Gaussian structure in the
senses given at the end of Sec.\ref{lang}, that is they produce non-trivial
non-Gaussian spectra. Here we shall find the most general type of isotropic
non-Gaussian structure which can be mapped from these theories.

We shall consider the anisotropic covariance matrix in more detail.
Let the matrix $C^{\ell\ell'}_m$ be split
into its diagonal and its off-diagonal $X^{\ell\ell'}_m$ parts
\begin{equation}
C^{\ell\ell'}_m=\delta^{\ell\ell'}C_{\ell|m|}+X^{\ell\ell'}_m
\end{equation}
Then $X^{\ell\ell'}_m\ll C_{\ell|m|}$, and so the bilinear form in
the exponent of the Gaussian distribution 
\begin{equation}
F(a^\ell_m)\propto\exp{\left(-{\sum_m}{\sum_{\ell,\ell'}}
a^\ell_m M^{\ell\ell'}_m a^{\ell'}_m \right)}
\end{equation}
is
\begin{equation}
M^{\ell\ell'}_m=C^{\ell\ell'-1}_m={\delta^{\ell\ell'}\over C_{\ell|m|}}
-{X^{\ell\ell'}_m\over C_\ell C_{\ell'}}
\end{equation}
and so the distribution factorizes into a factor which reveals the
structure inside each multipole, and a factor which reveals correlations 
between different multipoles. We shall analyze these two factors in turn.

Let's first assume that $X^{\ell\ell'}_m=0$.
Repeating the transformation presented in 
Section~\ref{lang} but using a covariance matrix of the form
($\ref{ani2p}$) one ends up with a rather complex distribution
which has the form:
\begin{equation}
F(C^\ell,B^\ell_m,s^\ell_0,\phi^\ell_m)=F(C^\ell,B^\ell_m)
\times {1\over 2}\times {1\over (2\pi)^\ell}
\end{equation}
Unless $C_{\ell m}=C_\ell$, the $B^\ell_m$ are not uniformly distributed.
Also the $C^\ell$ will in general not be a $\chi^2_{2\ell +1}$, and
the function $F(C^\ell,B^\ell_m)$ will not factorize. This means that
not only will correlations exist between the $B^\ell_m$ but the $B^\ell_m$
will also be correlated with the angular power spectrum. The phases
$\phi^\ell_m$ on the other hand will still be uniformly distributed
and independent. The phases tell us nothing about Gaussian anisotropic
fluctuations.

Hence anisotropic Gaussian fluctuations, when seen
from the point of view of an isotropic formalism, are an example  
of delocalized shapeful fluctuations (explored in some detail in 
\cite{fermag}). In the next two sections we will explore in more
detail the particular type of non-Gaussian effects which Gaussian
anisotropic fluctuations may induce. The shapes exhibited by these
theories are not the most general shapes, because there must be
a scale transformation in the $\rho^\ell_m$ which would render
the $B^\ell_m$ uniformly distributed again. Clearly not all shapes 
have this property. 

On top of this if $X^{\ell\ell'}_m\neq0$ the distribution $F(a^\ell_m)$
does not factorize into factors which only depend on one $\ell$.
Correlations between the different $\ell$ will then appear, which
in the language of \cite{fermag} amount to the emergence of connected
structures: different scales are allowed to interfere constructively.
In this paper we will not explore this side of the 
problem in depth. Nevertheless we have identified the non-Gaussian structures
into which anisotropic Gaussian fluctuations are mapped. These are
the delocalized shapeful (and possibly connected) structures 
defined in \cite{fermag}, or rather, a subclass thereof.

We should note that although the $\{C^\ell,B^\ell_m,\phi^\ell_m\}$
decomposition is not $SO(3)$ invariant, the $\{C^\ell,B^\ell_m\}$
already are $SO(2)$ invariant
\footnote{We are assuming that not only the Universe is anisotropic
but that we know, a priori, what its symmetry axis is, eg: by the
detection of a Hubble-size coherent magnetic field. Alternatively
we leave the Euler angles of this axis free, to be estimated by
some MLE.}.  
Since the phases contain no information whatsoever on Gaussian anisotropic
fluctuations they do not count as a device for making predictions
in these theories (as much as one does not compute $B^\ell_m$ for
Gaussian isotropic theories). Hence the set of variables 
$\{C^\ell,B^\ell_m\}$ is suitable for representing invariantly 
the most general form of non-Gaussian fluctuation which can be
mapped from Gaussian anisotropic fluctuations.

\section{Globally Anisotropic Universes}\label{glob2}
 A useful set of models in which to explore these concepts are the
homogeneous, anisotropic cosmologies, also know as the Bianchi models
\cite{bianchi}.
One can describe Bianchi cosmologies in terms of the metric
\begin{eqnarray}
g_{\mu\nu}=-n_\mu n_\nu+a^2[\exp(2\beta)]_{AB}e^A_\mu e^B_\nu,
\end{eqnarray}
where $n_\alpha$ is the normal to spatial hypersurfaces of
homogeneity, $a$ is the conformal scale factor, $\beta_{AB}$ is a 
$3 \time 3$ matrix only dependent on
cosmic time, $t$, and $e^A_\mu$ are invariant covector
fields on the surfaces of homogeneity, which obey the commutation
relations
\begin{eqnarray}
e^A_{\mu;\nu}-e^A_{\nu;\mu}=C^A_{BC}e^B_{\mu}e^C_{\nu}.
\end{eqnarray}
The structure constants
$C^A_{BC}$ can be used to classify the different models.
We shall focus on open or flat models which are asymptotically
Friedman. These can be obtained by taking different limits
of the type VII$_h$ model which has structure constants
\begin{eqnarray}
C^2_{31}=C^3_{21}=1,\  C^2_{21}=C^3_{31}=\sqrt{h}
\end{eqnarray}
It is convenient to define the parameter
$x=\sqrt{{h/{(1-\Omega_0)}}}$, which determines the scale on which
the principal axes of shear and rotation change orientation.
By taking combinations of limits of $\Omega$ and $x$ one can 
obtain Bianchi I, V and VII$_0$ cosmologies. 

We are interested in large-scale anisotropies so it suffices
to evaluate the peculiar redshift a photon will feel from
the epoch of last scattering (${\it ls}$) until now (0)
\begin{eqnarray}
{\Delta T_A}({\bf {\hat r}})=({\hat r}^iu_i)_{0}
-({\hat r}^iu_i)_{\it ls}-\int^{0}_{\it ls}{\hat r}^j{\hat r}^k
\sigma_{jk}d\tau 
\label{eq:dt1}
\end{eqnarray}
where ${\bf {\hat r}}=
(\cos\theta\sin\phi,\sin\theta\sin\phi,\cos\phi)$ 
is the direction vector of the incoming null
geodesic, ${\bf u}$ is the spatial part of the fluid four-velocity
vector and to first order, the shear is 
$\sigma_{ij}={\partial_{\tau}} \beta_{ij}$. To evaluate expression 
(\ref{eq:dt1}),
one must first of all determine a parameterization of geodesics
on this spacetime. This is given by
\begin{eqnarray}
\tan({{\phi(\tau)} \over 2}) &=&\tan({\phi_0\over 2})
\exp[-(\tau-\tau_0)\sqrt{h}] \nonumber \\
\theta(\tau)&=&\theta_0+(\tau-\tau_0) \nonumber \\-
{1 \over \sqrt{h}}
\ln\{\sin^2({\phi_0\over 2})&+&\cos^2({\phi_0\over 2})
\exp[2(\tau-\tau_0)\sqrt{h}]\}
\end{eqnarray}
Solving Einstein's equations (and assuming that matter is
a pressureless fluid)
one can determine ${\bf u}$ and $\sigma_{ij}$.
A general expression for (\ref{eq:dt1}) was determined in 
\cite{BJS}:
\begin{eqnarray}
{\Delta T}_A({\bf {\hat r}})&=&({\sigma \over H})_0{2\sqrt{1-\Omega_o} 
\over {\Omega_0}} 
\times \{ [\sin\phi_0\cos \theta_0-\sin \phi_{ls}
\cos \theta_{ls}(1+z_{ls})] \nonumber \\
&-& \int_{\tau_{ls}}^{\tau_0}{{3h(1-\Omega_0)} \over \Omega_0}
\sin 2\phi [\cos(\theta)+\sin(\theta)]{{d\tau} \over {\sinh^4 (
\sqrt{h}\tau/2)}}\} 
\label{eq:defTA}
\end{eqnarray}

A useful discussion of the different CMB patterns imprinted by 
 the unperturbed anisotropic 
expansion is presented
in \cite{BJS}. The patterns can be roughly said to be constructed
out of two ingredients: a focusing of the quadrupole when $\Omega<1$
and a spiral pattern when $x$ is finite. 
The Bianchi VII$_h$ is most general form of homogeneous, anisotropic universes
in an $\Omega\leq 1$ which are asymptotically
Friedman-Robertson-Walker.
The  pattern is of the form:
\begin{equation}
{\Delta T\over T} =f_1(\theta)\cos(\phi-{\tilde\phi}(\theta))
\end{equation}
In each $\theta={\rm const}$ circle the pattern has a 
dependence in $\phi$ of the form $\cos(\phi-{\tilde\phi})$. The
phase ${\tilde\phi}$ depends on $\theta$, and hence the spiralling of the
simple cold and hot bump induced by the $\cos\phi$ dependence.
The functions $f_1(\theta)$ and ${\tilde\phi}(\theta)$ are  rather 
complicated functions which have to be evaluated numerically, and depend
on various details of the particular Bianchi model within the type 
we have chosen. It is curious to note, however, that only the power
spectrum $C_\ell$ and the phases $\phi$ are sensitive to these details.
All the spirals imprinted by Bianchi VII$_h$ models have moduli
of the form
\begin{equation}
\rho^\ell_m=\delta_{m 1}f_2(\ell,x)
\end{equation}
Therefore their shape spectra will always be 
\begin{eqnarray}\label{blmspiral}
B^\ell_m&=&1\qquad {\rm for} \quad 2\le m\le \ell\nonumber\\
B^\ell_m&=&0\qquad {\rm for} \quad m=1
\end{eqnarray}
The background patterns in Bianchi VII$_h$ models are all localized, shapeful,
and connected structures. Depending on the model they will however have
different positions, power spectra, and connectivity. Nevertheless, 
their shape spectra is always the same exact shape, of form (\ref{blmspiral}),
without any cosmic variance error bars. Confusion with a Gaussian is zero.
Confusion with the shape of a perfect texture hot spot is zero as well.
These have a non-Gaussian spectrum of the form 
\begin{eqnarray}\label{blmspot}
B^\ell_m&=&1\qquad {\rm for} \quad 1\le m\le \ell\nonumber\\
\end{eqnarray}
Although the shape spectrum is the same up to the last $B^\ell_m$,
the confusion between the two theories  is zero.

\section{An Anisotropic Grischuk-Zel'Dovich effect}\label{gz}

We shall now consider an example of a flat homogeneous and isotropic
universe with topological identification along one axis. This example
is simpler than most of those considered in the literature but 
illustrates one of the key features
of such models: the breaking of statistical isotropy in the fluctuations.
Let us consider a universe with a topological identification along
the $z$ axis. All functions defined on such a  space satisfy:

\begin{eqnarray}
{\Phi}(x,y,z)={\Phi}(x,y,z+L) 
\end{eqnarray}

By considering a flat universe we can restrict ourselves to 
calculating the Sachs-Wolfe effect. The temperature anisotropy
from the  surface of last scattering will be given be:

\begin{eqnarray}
{{\Delta T}\over{T}}({\bf n})=-{1 \over 2}{H^2_0 \over c^2}\int d^2k
\sum_{j=-\infty}^{+\infty} 
\delta_{{\bf k}j}\frac{e^{i\Delta{\bf n}\cdot {\bf q}}}{q^2}
\end{eqnarray}
where $\Delta=\eta_0-\eta_{ls}$ is radius of the surface of last
scatter and $q=(k\cos\phi,k\sin \phi,2\pi \frac{j}{L})$. We can expand
the exponential in spherical harmonics to get

The $a_{\ell m}$s are given by
\begin{eqnarray}
a_{\ell m}&=&
-{i^{-\ell} \over 2}{H_0^2 \over c^2}\int d^2k\sum_{j}\delta_{{\bf k}j}
\frac{j_{\ell}(\Delta q)}{q^2}Y_{\ell m}^*({\hat q})
\end{eqnarray}

We now assume statistical homogeneity and isotropy of $\delta$ and
a scale invariant power spectrum
\begin{eqnarray}
\langle \delta^*_{{\bf k}j}\delta_{{\bf k'}j'}\rangle=\delta^2({\bf
k}-{\bf k'})\delta_{jj'}q^{-1} \ \ \ q=\sqrt{k^2+({{2\pi j}\over L})^2}
\end{eqnarray}
This leads
to the covariance matrix for the $a_{\ell m}$s:
\begin{eqnarray}
\langle a_{\ell m}^*a_{\ell 'm'}\rangle={i^{\ell'-\ell} \over 4}{H_0^4 \over
c^4}\int d^2k\sum_{j}\frac{j_{\ell}(q\Delta) j_{\ell'}(q\Delta) }
{q^3}Y^*_{\ell m}({\hat q})Y_{\ell' m'}({\hat q})
\end{eqnarray}

Expressing $Y_{lm}({\bf n})={\tilde P}_{lm}(\cos \theta)e^{im\phi}$
and performing the azimuthal integral, one immediately finds that
the covariance matrix is diagonal in $m$, so one has
\begin{eqnarray}
\langle a_{\ell m}^*a_{\ell 'm'}\rangle={i^{\ell'-\ell} \over 4}{H_0^4 \over
c^4}\delta_{mm'}\int kdk\sum_{j}\frac{j_{\ell}(q\Delta) j_{\ell'}(q\Delta) }
{q^3}{\tilde P}_{\ell m}({\hat q}){\tilde P}_{\ell' m'}({\hat q})
\end{eqnarray}
It is convenient to define $\chi=\Delta k$ and $\mu_j=2\pi\alpha j$
where $\alpha=\Delta/L$, i.e. the ration of our horizon to the
topological identification scale. If we now define $y=\mu_j/
\sqrt{\mu_j^2+\chi^2}$  the expression simplifies to
\begin{eqnarray}
\langle a_{\ell m}^*a_{\ell 'm'}\rangle={i^{\ell'-\ell} \over 4}\frac{H_0^4 \Delta}
{c^4}\delta_{mm'}\int_0^1 dy\sum_{j}\frac{j_{\ell}(\frac{\mu_j}{y}) 
j_{\ell'}(\frac{\mu_j}{y}) }
{\mu_j}{\tilde P}_{\ell m}(y){\tilde P}_{\ell' m'}(y)
\end{eqnarray}

In the limit where the identification  scale goes to infinity we
get the standard result
\begin{eqnarray}
\lim_{\alpha\rightarrow 0}\langle a_{\ell m}^*a_{\ell 'm'}\rangle\propto
\int_0^1dy j^2_{\ell}(\frac{1}{y})\delta_{\ell\ell'}\delta_{mm'}\propto
\frac{1}{\ell^2}\delta_{\ell\ell'}\delta_{mm'}
\end{eqnarray}
i.e. a scale invariant, diagonal covariance matrix. In the
case of finite $\alpha$ this is not the case. Consider the
quadrupole. The ring spectra has two components, $B_1$ and $B_2$
with a probability distribution function
\begin{eqnarray}
F(r,B_1,B_2)=\frac{\exp({-r^2/2\sigma^2_2)}r^4}{(\pi/2)^{1/2}\sigma_2^{3}3!!}
\times \exp\{-(r^2/2\sigma^2_2)[c_2B_2^2(1+\frac{c_1}{c_2}B_1^{2/3})]\}
\end{eqnarray}
where we have defined
\begin{eqnarray}
\sigma^2_2&=&(\langle |a_{20}|^2\rangle\langle |a_{21}|^2\rangle
\langle |a_{22}|^2\rangle)^{1/3} \nonumber \\
c_i&=&\frac{\sigma^2_2}{\langle |a_{2i-1}|^2\rangle}-
\frac{\sigma^2_2}{\langle |a_{2i}|^2\rangle}
\end{eqnarray}
By exploring the dependence of $c_1$ and $c_2$ on $\alpha$ 
we can see how the probability distribution function of the 
$B$s change with topology scale; to a very
good approximation we find 
\begin{eqnarray}
\frac{c_1}{c_2}&=& \frac{\sqrt{2}}{2} \nonumber \\
c_2&=& \frac{2\alpha}{1+\alpha} 
\end{eqnarray}
We can see the signature for non-Gaussianity arising here. For
a non zero $\alpha$ there are correlations between the
three statistical quantities, in particular
$\langle B_1B_2\rangle \propto \alpha/(1+\alpha)$. 
The probability distribution  function for $B_1$ and $B_2$ 
(defined on $[0,1]\times[0,1]$) becomes peaked at 0.
We have focused on the quadrupole where the effect is easy to
see. The method is systematic however, and one can construct the
probability distribution function of the high order ring spectra
in the same way.

This a curious application of the idea put forward in \cite{GZ}, an
anisotropic Grischuk-Zel'Dovich effect. By looking at the shape of
the low $\ell$ multipole moments we can constrain the degree of
statistical anisotropy outside the current horizon. Note that, 
already for $\alpha<1$ there are deformations in the covariance
matrix which may be statistically significant with current data.
This will be pursued in a future publication\cite{gzour}.

\section{Conclusions}\label{conc}
In this paper we have presented a new technique for quantifying
non-Gaussianity on large scales. It is the extension of the
non-Gaussian spectra developed in \cite{fermag} to the surface
of the two sphere. As we have shown in Section~\ref{lang} the
construction is slightly different to take into account the
particularities of the spherical harmonic basis. However the
qualititative interpretation of the different levels of
non-Gaussianity follows through, exactly as in \cite{fermag}.
One can identify the information contained in the ring, interring
and phase spectra with shape, connectivity and localization.

An interesting and untapped application is to universes with
statistically anisotropic fluctuations. Developing the idea
put forward in \cite{conf} we explain how statistical anisotropy
and non-Gaussianity are intimately related. From this one
can infer some novel properties of the covariance matrix
of fluctuations in statistically anisotropic spacetimes. In
particular, features which appear in non-gaussian theories of
structure formation, like textures \cite{ltx,ltx1} will appear
here: a surplus of cosmic variance and cosmic {\it co}-variance
of the power spectra.

There are a number of situations where these results are applicable.
One is in the case of anisotropic universes, i.e. universes
which aren't Friedman-Robertson-Walker. There are a number of
known examples \cite{bianchi,krasinski}. Without actually doing 
perturbation theory on them we argue for a natural prescription
for adding fluctuations in the CMB to such models. It consists
of finding the reduced symmetry group of the temperature patterns
and constructing anisotropic Gaussian random fields with
such properties (in which the covariance matrix satisfies those
symmetries, and not more). In fact it can be shown that these
symmetries can be deduced from the geodesic structure of
the space time \cite{joja}. This can be a first step in extending
the prescription used in \cite{bfs} for constraining general
anisotropic models with the COBE 4 year data. A brief analysis
is made of the relevant Bianchi models for which we present the
non-Gaussian spectra.

Another, different application is  the case of homogeneous
isotropic models where an anisotropic topological identification
has been imposed. As an example we identify one direction in
space. One finds that statistical isotropy is broken. This can be
easily from the following: if we look along the axis of
identification, and the identification scale is smaller than our horizon,
one will find strong correlations between patches of the
microwave sky which are reflected about the uncompactified plane\cite{CSS}.
By looking at the structure of the covariance matrix one can see
that this anisotropy will manifest itself by inducing not only
non-Gaussian ring spectra but also inter-ring spectra. This
non-Gaussian manifestation may persist if we consider the
identification scale to be large than our horizon. We name
this effect the {\it anisotropic} Grischuk-Zel'Dovich effect.

Although we now have a high quality measurement of anisotropies
on large angular scales we are confronted with the hardships of
the real world. Galactic contamination leads one to consider an
anisotropic rendition of the sky and considerably complicates
the analysis of the COBE four year data. It is well known that
one of the consequences is that the quadrupole measurement should
be viewed with scepticism. Unfortunately it is the quadrupole
which could supply us with a probe of primordial CMB on the
largest angular scales. There may be ways around these shortcomings.
One can try and reformulate our non-Gaussian spectra on the largest
angular scales using the techniques put forward in \cite{gorski}.
This would involve a proper likelihood analysis and to make
the problem tractable one would need to find an adequate
parametrization of  the non-Gaussian spectra. The fact that
we have devised a consistent method for characterizing
non-Gaussianity on all scales may allow us to use the cumulative
information of all $\ell>2$ to infer the behaviour on large scales;
all modes will be affected to some extent by large scale anisotropy.
Finally it would be interesting to analyse in more detail the
observability of the anisotropic Grischuk-Zel'Dovich effect,
taking into consideration issues of cosmic variance.

\section*{Acknowledgements}
We would like to thank T Kibble for pointing out and correcting
a serious mistake at the heart of this paper. J.M. thanks MRAO-Cambridge
for use of computer facilities while this paper was being prepared.
P.G.F gratefully acknowledges insightful comments from C. Balland, J. Levin,
A. Jaffe and J. Silk.
P.G.F. was supported by  the
Center for Particle Astrophysics, a NSF Science and
Technology Center at UC Berkeley, under Cooperative
Agreement No. AST 9120005. J.M. was supported by a Royal
Society University Research Fellowship.

\section*{Appendix I - Gaussian fluctuations with a symmetry}
In this Appendix we detail the group theory argument sketched in
Section~\ref{glob1}. This argument is not necessary in the $SO(2)$ case
targeted in this paper, where the covariance matrix may be
easily derived directly. However, it opens doors to
more general anisotropic symmetry groups. It also shows how
the general form of the covariance matrix depends not on the
symmetry group, but only on its irreducible representations (Irreps).

Fluctuations (Gaussian or not) in any Universe 
must be subject to the symmetries
of the underlying cosmological model. However, due to the random nature
of the fluctuations, they must satisfy these symmetries only statistically.
By this we mean that the statistical ensemble of fluctuations,
and not each realization, should be subject to the symmetries.
If a symmetry transformation is applied to each member of the ensemble,
then each member may change, but the ensemble should remain the same. 
For instance, for the $SO(3)$ symmetry group only realizations 
containing only a monopole $\ell=0$ are left unchanged by rotations.
Nevertheless much more general fluctuations respect statistical isotropy.
Similarly, only the $m=0$ modes are cylindrically symmetric, 
but more general fluctuations are $SO(2)$ statistically 
invariant\footnote{Some texture models exhibit approximate $SO(2)$ symmetry
at low $\ell$ in {\it each} realization. Then an axes system exists in which
the $m=0$ mode has much more power than any other.
This is of course a very non-Gaussian
effect which cannot be simply reproduced by anisotropic Gaussian
fluctuations.}.

Gaussian fluctuations are fully specified by their covariance matrix
\begin{equation}
C^{\ell_1\ell_2}_{m_1m_2}={\langle a^{\ell_1}_{m_1}a^{\ell_2\star}_{m_2}
\rangle}
\end{equation}
which may be seen as a bilinear form on the $\{a^\ell_m\}$ space.
Hence the statistical symmetries of a Gaussian theory are equivalent
to the requirement that the covariance matrix is left unchanged by any
symmetry transformation. Let $G$ be the symmetry group of the underlying
cosmological model as projected on the sky. Let's first suppose that 
$G$ breaks the $\{a^\ell_m\}$ into a set on non-equivalent Irreps.
Then, let's find a new basis $a^L_M$ adapted to $G$, where $L$ now 
labels the Irrep the basis element belongs to, and $M$ the actual
element. Then $G$ is represented by a set of matrices $G^L_{MM'}$
acting on $a^L_M$ as
\begin{equation}\label{transf}
{\tilde a}^L_M=Ga^L_M=G^L_{MM'}a^L_{M'}
\end{equation}
The covariance matrix for the $a^L_M$ 
\begin{equation}
C^{L_1L_2}_{M_1M_2}={\langle a^{L_1}_{M_1}a^{L_2\star}_{M_2}\rangle}
\end{equation}
must remain unchanged by the transformation (\ref{transf}), so that
\begin{equation}
{\tilde C}^{L_1L_2}_{M_1M_2}={\langle {\tilde a}^{L_1}_{M_1}
{\tilde a}^{L_2\star}_{M_2}\rangle}= G^{L_1}_{M_1M'_1}G^{L_2\star}
_{M_2M'_2}C^{L_1L_2}_{M'_1M'_2}=C^{L_1L_2}_{M_1M_2}
\end{equation}
which for unitary representations amounts to the commutation
relation
\begin{equation}
G^{L_1}C^{L_1L_2}=C^{L_1L_2}G^{L_2}
\end{equation}
Let us now recall Schur's Lemmas \cite{cornwell}.
\newtheorem{theorem}{Schur's Lemma}
\begin{theorem}
Let $\Gamma$ and $\Gamma'$ be two Irreps of a group $G$ with dimensions
$d$ and $d'$, and let there be a $d\times d'$ matrix $A$ such that
\begin{equation}
\Gamma(g) A=A\Gamma'(g)
\end{equation}
for all group elements $g$. Then either $A=0$ or $d=d'$ and 
${\rm det}A\ne 0$.
\end{theorem}
It follows that if $A\ne 0$ then $\Gamma$ and $\Gamma'$
are equivalent.
\begin{theorem}
If $\Gamma$ is a $d$-dimensional Irrep of a group $G$ and $B$ is a
$d\times d$ matrix such that 
\begin{equation}
\Gamma(g) B=B\Gamma(g)
\end{equation}
for all group elements $g$, then $B=\lambda 1$.
\end{theorem}
Combining the two Lemmas we can then find that the covariance matrix
must take the form
\begin{equation}
C^{L_1L_2}_{M_1M_2}={\langle {\tilde a}^{L_1}_{M_1}
{\tilde a}^{L_2\star}_{M_2}\rangle} =\delta^{L_1L_2}\delta_{M_1M_2}C_{L_1}
\end{equation}
An interesting result is that if $G$ breaks the $\{a^\ell_m\}$ space into 
non-equivalent Irreps spanned by  $\{a^L_M\}$, then these must be independent 
random variables with a variance which can only depend on the Irrep
they belong to. This argument applies for instance if
the symmetry group is $SO(3)$, in which case $L=\ell$ and $M=m$.

The argument just present breaks down however, if some of the Irreps
defined by $G$ are equivalent. Let the $\{a^\ell_m\}$ space now be spanned
by an adapted basis $\{a^{LD}_M\}$, where $L$ labels each class of equivalent
representations, $D$ labels the actual Irrep within this class, 
and $M$ labels the elements
within each Irrep. Then, from the second Schur's Lemma we know that
within the same Irrep we still have:
\begin{equation}
{\langle a^{LD}_{M_1} a^{LD\star}_{M_2}\rangle}=\delta_{M_1M_2}C^{LD}
\end{equation}
but for different but equivalent Irreps ($D_1\ne D_2$) we now have
\begin{equation}
{\langle a^{LD_1}_{M_1} a^{LD_2\star}_{M_2}\rangle}=C^{LD_1D_2}_{M_1M_2}
\end{equation}
with ${\rm det } C\ne 0$, 
whereas for non-equivalent Irreps ($L_1\ne L_2$) we still have
\begin{equation}
{\langle a^{L_1D_1}_{M_1} a^{L_2D_2\star}_{M_2}\rangle}=0
\end{equation}
Therefore, although the $a^{LD}_M$ are independent within each Irrep
and among different non-equivalent Irreps, correlations may exist
between different but equivalent Irreps. Of course one may always
rotate the $a^{LD}_M$ within each class of equivalent Irreps
so as to diagonalize the covariance matrix. However such a rotation
is model dependent, and cannot be determined from the symmetries.
This situation happens for instance in the case of $SO(2)$, with
$L=m$ and $D=\ell$ (no $M$ index, since the Irreps are one dimensional).
Each $m$ provides a class of equivalent representations, with the
same $m$ but different $\ell$. 
Then the covariance matrix takes the general form
\begin{equation}
{\langle a^{\ell_1}_{m_1} a^{\ell_2\star}_{m_2}\rangle}
=\delta_{m_1m_2}C^{\ell_1\ell_2}_{m_1}
\end{equation}
and as we see although correlations among different $m$ are not
allowed, now we may have correlations between different $\ell$,
for the same $m$. We could rotate the $a^\ell_m$ in $\ell$ for
each fixed $m$ so as to diagonalize the covariance matrix, but
such procedure naturally would depend on the covariance matrix one
starts from, and would therefore be model dependent.


\begin{thebibliography}{99}
\bibitem{note} A simple example is supplied in \cite{ltx1}: an
inflationary theory which predicts a random tilt.
\end{thebibliography}
\end{document}